# Explanation of Faraday's Experiment

# by the Time-Space Model of Wave Propagation

Alexei Krouglov

*Sicon Video Corporation*

*1550 16th Ave., Bldg. F, 2nd Floor, Richmond Hill, Ontario L4B 3K9, Canada*

Email: Alexei.Krouglov@SiconVideo.com

# ABSTRACT

It is shown by the means of Time-Space Model of Wave Propagation the underlying phenomena of the alternating current's origin in famous Faraday's experiment.

The Time-Space Model of Wave Propagation is applied to the description of fluctuation phenomena both in physics and in economics [1, 2, 3].

In this paper the Model is applied to the Faraday's experiment of rotating a rectangular conducting loop in a magnetic field [4].

Magnetic field represents the energy's level in the Model terminology. We choose two points on the opposite sides of a loop (see **Fig. 1**) and calculate the induced values of energy's disturbances.

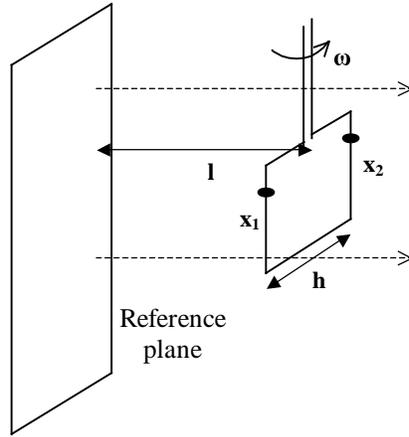

**Fig. 1**

Some other values are also shown on **Fig. 1** as side $h$ of the loop, distance $l$ from loop's center to the "reference plane", which is perpendicular to the energy's level gradient changes, and an angular speed $\omega$ of loop's rotation.

Then the distances from $x_1$ and $x_2$ to the reference plane are

$$x_i = l + \frac{h}{2}\sin(\omega t + \varphi_i), \qquad (1)$$

where $i = 1, 2$ and $\varphi_i$ are the initial phases, $\varphi_2 = \varphi_1 + \pi$.

I suppose that the energy's values in both points $x_1$ and $x_2$ are equal,

$$U(x_1,t) = U(x_2,t) \tag{2}$$

for $\forall t$.

I also assume that the energy's level at the reference plane is equal everywhere $\Phi(0,t)$ and is distinct from the energy's level at rest $\Phi_0$ on some value $\Delta\Phi_0$ i.e.

$$\Phi(0,t) = \Phi_0 + \Delta\Phi_0. \tag{3}$$

Therefore we can calculate the energy's levels in points $x_1$ and $x_2$ using [1],

$$\Phi(x_i,t) = \Phi_0 + \Delta\Phi_0 \cdot e^{-\mu x_i}, \tag{4}$$

where $\mu > 0$ is some constant.

Thus the energy's disturbances at points $x_1$ and $x_2$ are

$$\Delta U(x_i,t) = \Delta\Phi_0 \cdot e^{-\mu x_i}. \tag{5}$$

We can find the difference between the values of energy's disturbances in points $x_1$ and $x_2$ respectively,

$$\Delta U(x_1,t) - \Delta U(x_2,t) = 2 \cdot \Delta\Phi_0 \cdot e^{-\mu l} \sinh\left(-\mu \frac{h}{2} \sin(\omega t + \varphi_1)\right). \tag{6}$$

Changing in time difference between values of the energy's disturbances causes the energy's disturbance propagation, which is manifested in the form of alternating current in the loop.